# Aligning Three-Decade Surge in Urban Cooling with Global Warming

Haiwei Li [1, 2], Yongling Zhao [1, *], Ronita Bardhan [2, *], Pak-Wai Chan [3], Dominique Derome [4], Zhiwen Luo [5], Diana Ürge-Vorsatz [6], Jan Carmeliet [1]

[1] Department of Mechanical and Process Engineering, ETH Zürich, Zürich, Switzerland

[2] Department of Architecture, University of Cambridge, Cambridge, United Kingdom

[3] Hong Kong Observatory, Kowloon, Hong Kong, China

[4] Department of Civil and Building Engineering, Université de Sherbrooke, Sherbrooke, Canada

[5] Welsh School of Architecture, Cardiff University, Cardiff, Wales, United Kingdom

[6] Department of Environmental Sciences and Policy, Central European University, Austria

*Corresponding authors: Yongling Zhao (yozhao@ethz.ch), Ronita Bardhan (rb867@cam.ac.uk)

**Abstract**

Rising demand for space cooling has been placing enormous strain on various technological, environmental, and societal dimensions, resulting in issues related to energy consumption, environmental sustainability, health and well-being, affordability, and equity. Holistic approaches that combine energy efficiency optimization, policy-making, and societal adaptation must be rapidly promoted as viable, timely solutions. We interpret the 30-year upward trend and spikes in urban cooling demand from the perspective of climate change, urbanization, and background climates, focusing on five representative cities: Hong Kong, Sydney, Montreal, Zurich, and London. An unequivocal, worrying upward trend in cooling demand is observed in meteorological data from 1990 to 2021, using cooling degree hours (CDH) as a city-scale metric. The surge in cooling energy demand can be largely attributed to global warming, urban heat islands, and extreme heat events. Further, our quantification of the impact of the base temperature, in relation to the historical CDH, reveals that a 20% energy saving could be achieved instantly within a rather broad range of temperature and humidity by increasing the setpoint temperature by one degree, while characteristic sensational and physiological levels can be maintained at 'acceptable' and 'physiological thermal neutrality' respectively. However, the potential of reducing cooling demand can be nonlinearly and significantly lowered due to the presence of compound high relative humidity and high air temperature. To reduce cooling energy demand rapidly in a warming climate, we highlight the necessity of promoting hard and soft behavioral adaptation along with regulatory intervention for the operation of space cooling systems.

## 1. Introduction

The building sector is responsible for 30% of global energy consumption and 27% of total energy emissions (IEA 2022). Currently, 56% of the world's population lives in urban areas, and this percentage is expected to increase to 68% by 2050 (Desa 2018), escalating the energy burden in cities. The energy demand for space heating or cooling is determined by building design and operation, building physical properties and occupancy activities, urban design and availability of green and blue infrastructure, socio-economic development, and most importantly, climatic conditions (Biardeau *et al* 2020, Ürge-Vorsatz *et al* 2015, Zhao *et al* 2023). The earth's average surface temperature has risen by approximately 1.2 °C since the late 19[th] century. Meanwhile, the frequency, duration, and severity of extreme temperature events, such as heatwaves and record-breaking high temperatures, are aggravated (Coumou and Rahmstorf 2012). Due to the urban heat island (UHI) effect, these events occur more often in urban areas than in the surrounding suburban or rural areas (Zhao *et al* 2014). This phenomenon frequently appears alongside urbanization, with enlarged urban heat storage capacity, long-wave radiation trapping as a result of low reflectivity urban surfaces, reduced evapotranspiration and convection efficiency, and increased anthropogenic heat production. The UHI effect exacerbates the overheating in cities, leading to high heat-related illness and higher mortality rates, increased building cooling energy demand, air pollution associated with the high temperatures and reduced ventilation, and a potential energy crisis among a large portion of the global population (Li *et al* 2019). From June 2022 to August 2022, heatwaves in Europe were reported as the most deadly meteorological events and they have caused more than 20,000 heat-related deaths. The global air-conditioning demand is expected to increase rapidly, and it is expected that the electricity demand for cooling in 2100 will be 40 times greater than it was in 2000 (Isaac and van Vuuren 2009). Understanding the long-term surge of cooling energy demands and the causal factors is crucial for predicting future energy demand and related emissions and formulating the implementation of mitigation strategies of the Intergovernmental Panel on Climate Change (IPCC) and Sustainable Development Goals (SDGs).

For a 30-year period, we analyzed the cooling demand and the principal climatic drivers, such as warming background climate by global warming, urban heat island (UHI), heatwaves, and some potential mitigation measures for five cities residing in different climates: Hong Kong, Sydney,

Montreal, Zurich, and London. The cooling demand for urban and rural (or suburban) areas is quantified by yearly cooling degree hours (CDH) calculated from climatological standard 30-year (from 1990 to 2021) observation data.

## 2. Methods

*2.1 Study sites*

In order to investigate the cooling demand variation related to air temperature across regions with diverse climate types, urban morphology patterns, and levels of urbanization, we conducted an analysis of climatic data in urban and suburban/rural areas of five cities: Hong Kong, Sydney, Zurich, Montreal, and London. These cities were chosen due to their high levels of urbanization and population densification, making them common subjects of research in the field of UHI studies.

The selection of urban and suburban/rural sites was based on the local climate zone (LCZ) classification, a widely recognized standard in UHI research for classifying urban morphologies and natural landscapes. The LCZ classification system consists of ten built-up classes and seven land cover types (Lau *et al* 2019), providing a comprehensive parameterization of characteristics such as building height and coverage, pervious/impervious cover, aspect ratio, and surface materials (Núñez Peiró *et al* 2019). The suburban/rural sites were selected within a 40 km radius of the corresponding urban sites. The urban sites were chosen to represent a range of LCZ categories ranging from Compact to Open and High rise to Low rise (LCZ-1 to LCZ-6), while the suburban/rural sites were predominantly characterized by Trees and plants (LCZ-A to LCZ-D), Water (LCZ-G), or Bare soil or sand (LCZ-F).

Hong Kong falls within the monsoon-influenced humid subtropical climate zone (Köppen climate classification: Cwa), while Sydney is classified as humid subtropical (Cfa). Both cities experience hot and humid summers, along with cool to mild winters, with monthly mean temperatures exceeding 18 °C and less pronounced dry seasons. The urban site in Hong Kong is situated at the Observatory Headquarters (latitude, longitude: 22.30, 114.17), and the suburban site is located in Ta Kwu Ling (22.53, 114.16). In Sydney, the urban site is located in Bankstown (latitude, longitude: -33.92, 150.98), while the suburban site is situated in Observatory Hill (-33.86, 151.20).

Montreal belongs to the warm-summer humid continental climate zone (Dfb), characterized by four distinct seasons, significant temperature variations throughout the year, and moderately distributed precipitation. The urban site in Montreal is found in Trudeau (latitude, longitude: 45.47, -73.74), and the suburban site is located in Mirabel (45.68, -74.04).

Zurich exhibits a climate that lies between the warm-summer humid continental (Dfb) and temperate oceanic (Cfb) zones. Similarly, London falls within the temperate oceanic climate zone (Cfb). These cities experience cool summers and mild winters, with relatively narrow monthly mean temperature ranges and low seasonal variations. The urban site in Zurich is situated in Kaserne (latitude, longitude: 47.38, 8.53), while the rural site is located in Kloten (47.29, 8.32). In London, the urban site is located in St James' Park (latitude, longitude: 51.50, -0.23), and the rural site is situated in Kenley (51.30, -0.09).

*2.2 Meteorological data and cooling degree hour (CDH) calculation*

The analysis utilizes three decades' hourly air temperature data at the two-meter height at the selected weather stations. This parameter is the ambient temperature at two meters above surfaces of land, sea or water, which is valuable thermal information on outdoor pedestrian conditions (Stathopoulos 2006). The time period of data is subject to the data availability of the local weather stations, Hong Kong (urban and suburban: 1990 to 2021), Sydney (urban: 1993 to 2021, suburban: 1991 to 2021), Montreal (urban and rural: 1990 to 2021), Zurich (urban: 1991 to 2020, rural: 1990 to 2020), London (urban and rural: 1990 to 2021).

In order to quantify the air temperature-related cooling load pattern, this work employs cooling degree hour (CDH) calculation, a non-invasive measurement to analyze the accumulated cooling needs over a specific time period based on the climatic change of the outdoor environment. It provides energy consumption patterns in relation to weather conditions rather than exact values of cooling loads. CDH is a common tool for understanding the energy trends and extremes caused by climatic conditions and variations (Salata *et al* 2022, McGarity and Gorski 1984). And it is useful in comparative analysis over a long period of time among multiple areas and making potential cooling efficiency improvements based on comparative analysis. Detailed descriptions and discussions are presented in Appendix A.

**3. Results and discussion**

*3.1 Aligning cooling demand with background climates and urbanization*

Figure 1 (a) shows the results of CDH calculations for the five cities mentioned above from 1990 to 2021, demonstrating yearly cooling energy demand in all five cities during the last three decades. The base temperature in the CDH calculation is 22 °C. The main climate driver of energy demand is the ambient background temperature during the cooling season (van Ruijven *et al* 2019). The increasing trend has a robust association with the temperature, including increasing time-averaged temperature, increasing peak temperatures, and heatwave events during the cooling season. The background climatology determines the distinct difference in the magnitudes of the cooling demand in five cities.

The cooling season refers to the period that requires cooling to maintain the inside building temperature below the setpoint temperature. Both Hong Kong and Sydney have a cooling season from Spring to Autumn, while the cooling season for Montreal, Zurich, and London is mostly Summer, from June to August. Hong Kong shows a value of CDH in the range of 25 to 33 k°C·h and Sydney shows a CDH in the range of 2 to 10 k°C·h. Montreal and Zurich show CDH ranging from 1 to 7 k°C·h. Moreover, London has CDH values up to 3 k°C·h. The results of CDH show that the warmer subtropical cities, i.e., Hong Kong and Sydney, have distinctly higher cooling demand and longer cooling seasons than Montreal, Zurich, and London.

Due to the nature of CDH calculation and data availability, our analysis (Figure 1a) shows the cooling demand patterns associated with air temperature. In the actual setting, the population density (Figure 1b) and some other socio-economic factors, such as annual income and energy prices, may also contribute to the change in cooling demand. Spinoni et al. (2018) incorporated population weighting to analyze the socio-economic sensitivity. Since our focus is the climate-induced cooling load quantification and the socio-economic data is not available in all five cities, we did not incorporate socio-economic in CDH calculation. To complement the CDH calculation, we made an additional analysis regarding the trend of population density in all five cities. The continuous growth of population density indicates that the actual cooling demand of these cities is potentially higher than our reported values (Manoli *et al* 2019).

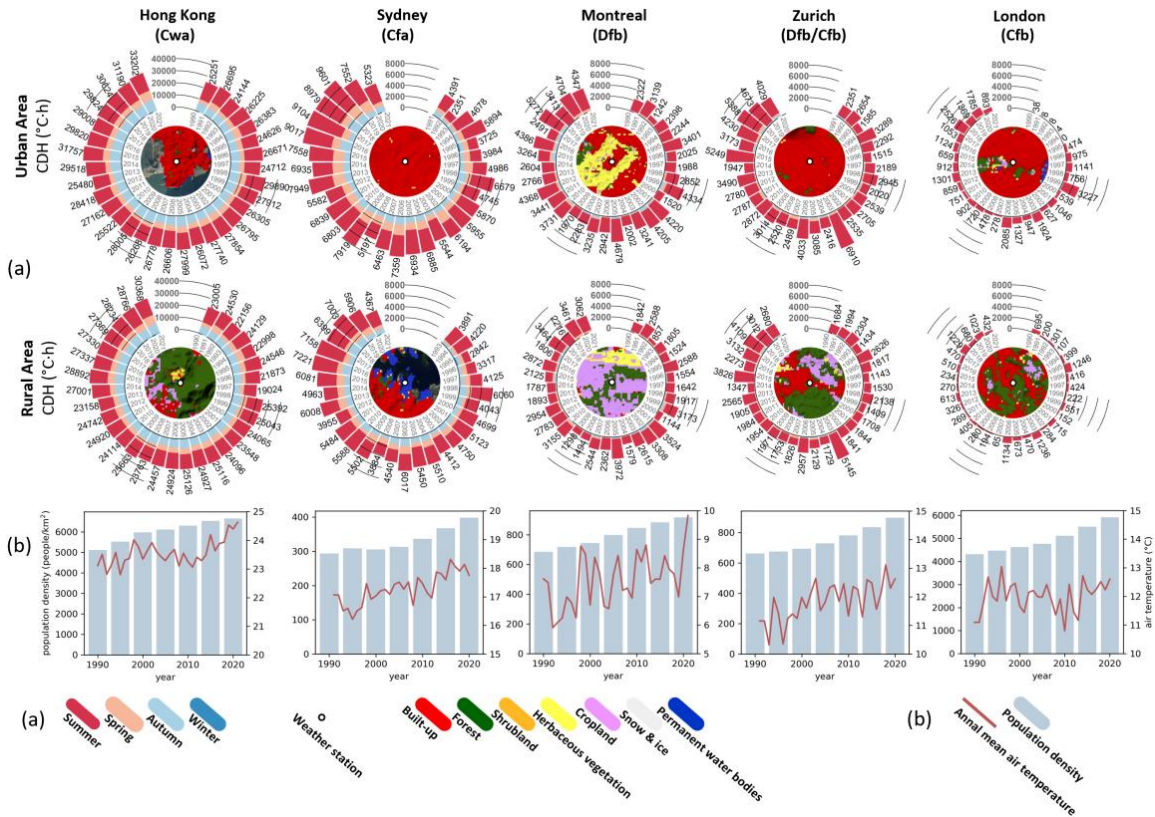

**Figure 1.** (a) Yearly cooling degree hours (CDH) of urban and rural areas (or suburban) in Hong Kong, Sydney, Montreal, Zurich, and London from 1990 to 2021. The base temperature for CDH calculation is 22 °C. The graphs in the center of each circular plot display the land types within 3 km of the weather stations. Remark the maximum scale for Hong Kong, 40,000 is different from the other cities with a maximum scale of 8,000. (b) Population growth (bars) and annual mean urban air temperature rise (lines) from 1990 to 2021. Note that the sub-figures have different scales of the y-axis for different cities.

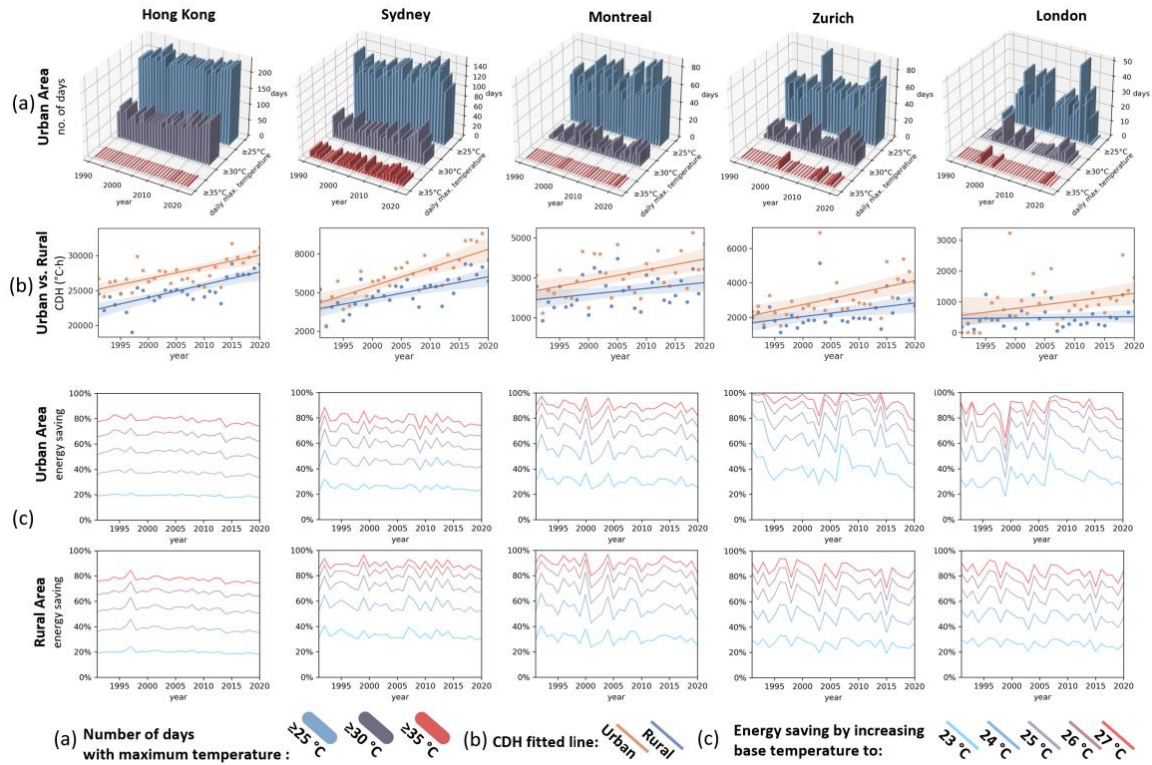

**Figure 2.** (a) The number of days with maximum air temperature ≥25, 30, 35 °C in the urban areas of Hong Kong, Sydney, Montreal, Zurich, and London, based on the calculation of climate norms stated in WMO 2017 Guidelines (Arguez and Vose 2011). (b) Yearly urban and rural CDH with fitted trend-lines (orange: urban, blue: rural/suburban). The CDH data are calculated based on base temperature, i.e., 22 °C. (c) Energy saving or CDH reduction by increasing the base temperature from 22 °C to 23-27 °C. The increase in the base temperature reflects the potential impact of behavioral adaptation and regulatory intervention for spacing cooling.

*3.2 Cooling demand spikes in extreme heat events and heatwaves*

High-temperature events can be distinguished by the number of days with the maximum temperature exceeding 25 °C, 30 °C, and 35 °C, shown in Figure 2 (a), as proposed in the WMO Guidelines. These events also can be seen in the yearly CDH, where spikes in CDH can be interpreted as indicators of extreme heat events in the considered year. Extreme heat events contribute to more than half percentage of CDH. More specifically, CDH in the heatwave period in 2022 takes up above 71% (urban Hong Kong), 60% (urban Sydney), 46% (urban Montreal), 44% (urban Zurich), and 47% (urban London) of the total CDH for the whole year. The cooling load is more than doubled in the years with exceptionally high frequency and high duration of summer heat events. The increasing trends of CDH are relatively smooth in Hong Kong, Sydney, and Montreal, while the spikes in CDH is more frequently seen in western Europe, e.g., London and Zurich. Rousi et al. (2022) have identified Europe as a 'heatwave spot' since its increase in the occurrences of extreme heat has been three-to-four times more frequent than for other northern mid-latitude regions over the past decades, which is mainly due to the increasing trend in the persistence of double jet occurrences. Zurich and London suffered from the record-breaking heatwave that prevailed in Europe in the year 2003. That severe heatwave is still considered to be the warmest period of the last 500 years, which not only caused energy consumption to increase but also burdened health and emergency services in Europe, leading to over tens of thousands of excess deaths. Recently, an exceptional heatwave event affected the U.K. in July 2022, reaching 40 °C for the first time and causing over 2,800 excess deaths in the elder population. In Switzerland, MeteoSwiss activated orange and yellow alerts for heatwaves in 2022 and recorded a temperature up to 38.3 °C in August 2022. Urgent and effective climate-sensitive urban planning with sustainable and resilient mitigation measures is critical to tackling future energy demand spikes.

*3.3 Three-decade cooling demand in global warming*

In Figure 2 (b), the mean trend-lines of CDH show an approximate increase rate of 160 °C·h per year (170 °C·h per year) over the last 30 years of urban (suburban) Hong Kong, 120 °C·h per year (70 °C·h per year) for urban (suburban) Sydney, 50 °C·h per year (20 °C·h per year) for urban (rural) Montreal, 30 °C·h per year (25 °C·h per year) for urban (rural) Zurich and 25 °C·h per year (5 °C·h per year) for urban (rural) London. With respect to 1990, the CDH increased by 20% (23%) for urban (suburban) Hong Kong, 100% (83%) for urban (suburban) Sydney, 60% (50%) for urban (rural) Montreal, 100% (65%) for urban (rural) Zurich and 160% (30%) for urban (rural) London. The trend-lines show that, for all cities except with Hong Kong, the relative increase in CDH is higher in urban regions compared to rural regions, referring to the increasing UHI effect with rising temperatures due to global climate change. The reduction of evapotranspiration and convection efficiency and the increase of anthropogenic heat in urban areas are considered the main contributors to urban overheating (Krayenhoff *et al* 2021, Zhao *et al* 2021). As presented in Figure 1 (a), urban areas have more impervious heat-storing built-ups and less vegetation or water bodies than rural areas, meaning low water availability and evapotranspiration in urban environments, leading to high urban-rural temperature differences and higher cooling demand in urban areas (Zhao *et al* 2023, Yang *et al* 2022). The convection efficiency, which is associated with changes in the aerodynamic resistance, represents the heat transfer from the surfaces of buildings to the atmosphere (Zhao *et al* 2014). The high aerodynamic resistance of urban areas results in low efficient convection, which reduces the convection efficiency and increases the UHI intensity and cooling demand. Re-introducing green spaces and water surfaces into the urban area could increase both evapotranspiration and convection efficiency, and effectively reduce the energy demand for cooling.

*3.4 Energy saving potential by behavioral adaptation and regulatory intervention*

As the base temperature is increased to 23-27 °C, energy saving or CDH reduction with respect to that at the base temperature of 22 °C is dramatic (Figure 2c). The base temperature is a fundamental consideration in CDH analysis, above which cooling is required, and it should be determined based on local climate, occupancy activities, building properties and functions, air conditioning systems and cultural factors. Prior research used a common base temperature of around 22 °C for CDH calculations (Bolattürk 2008). Upgrading building cooling systems using an adaptive comfort approach and higher thermostat setpoints, such as the Cool Biz approach in Japan (Murakami *et al* 2009), can increase the base temperature for indoor space cooling. The cooling demand can thus be approximately reduced by 20% as the base temperature is merely increased by one degree, i.e., from 22 °C to 23 °C, implying that behavioral adaptation may make an immediate difference in achieving desired energy saving without compromising thermal comfort considerably. Although setpoint air temperature is usually considered the most crucial parameter in determining thermal comfort in an indoor environment, behavioral adaptation for other setpoint parameters, such as the indoor humidity level, is also necessary for maintaining acceptable adaptive comfort levels in an indoor environment (ASHRAE 2005).

High humidity levels can hinder sweating and thus evaporative cooling from skin, making individuals feel hotter and potentially uncomfortable, which can cause extra discomfort or even dangerous heat stress under heat waves and extreme weather (Zhang *et al* 2023). Here, we explore the impact of relative humidity on indoor thermal comfort using the Standard Effective Temperature (SET) as a metric where the SET measures perceived temperature and physiological response of individuals by combining various environmental parameters in a given environment (Tartarini *et al* 2020, Zhang and Lin 2020, Ji *et al* 2022, Arguez and Vose 2011). As presented in Appendix B, it is found that by adjusting indoor relative humidity while maintaining a fixed setpoint of air temperature, the variations in SET values could be up to 1.7 °C. When the setpoint of air temperature is set at 26°C, increasing the setpoint of relative humidity from 40% to 50% can alter a person feeling from neutral to slightly warm. Therefore, in order to achieve energy-saving while maintaining a comfortable thermal comfort level in the environment, it is possible to increase the base temperature to a certain level while carefully considering setpoints of other parameters. The potential for reducing CDH, while maintaining the sensational and physiological levels, can be nonlinearly and significantly affected by the level of indoor relative humidity. The presence of compound high relative humidity and high air temperature makes it challenging to reduce CDH effectively.

Renovating existing building systems to improve energy performance is crucial worldwide, as already pointed out by the Commercial Building Disclosure (CBD) program in Australia and in the Annex projects launched by the International Energy Agency (IEA). With implementation of passive cooling and thermal retrofitting for buildings, higher setpoints of temperatures up to 27 - 28 °C could be adopted for operation of space cooling. Effective regulatory interventions on building energy retrofitting and operation codes should be preferred instruments for policymakers aiming to achieve rapid reduction in cooling energy demand.

## 4. Conclusion

Our research, starting with the analysis of meteorological data in five representative cities, Hong Kong, Sydney, Montreal, Zurich, and London, reveals a significant upward trend in cooling demand over the past three decades. The background climates, impervious built-up surfaces, population density, and some socio-economic factors largely influence the cooling demand in cities. Additionally, extreme heat events and heatwave events can result in more than doubled cooling spikes in urban areas, which are evident in heatwaves occurred in European cities (London and Zurich). An appropriate base temperature selection should be based on a thorough consideration of setpoints of multiple parameters for thermal comfort rather than the air temperature alone. The quantification of the impact of base temperature on cooling degree hours (CDH) indicates that a one-degree increase in setpoint temperature, with behavioral adaptation or regulatory intervention on the operation of space cooling systems, could result in 20% energy savings while sensational and physiological characteristics can be maintained. The potential for reducing CDH diminishes dramatically when both relative humidity and air temperature are at high levels. Our findings provide insights into developing sustainable, viable, and timely solutions to mitigate the surge of cooling demand in the time of global warming.

**Acknowledgment**

The authors wish to acknowledge the meteorological data provided by the Australian Bureau of Meteorology and Hong Kong Observatory.

**Data availability**

A preview of the CDH data is available via https://github.com/florahww/Urban-Cooling-Data.git. The ambient temperatures and CDH dataset will be accessible after publication both on GitHub and on the website of the Chair of Building Physics, ETH Zurich.

**Acknowledgment**

The authors wish to acknowledge the meteorological data provided by the Australian Bureau of Meteorology and Hong Kong Observatory.

**Data availability**

A preview of the CDH data is available via https://github.com/florahww/Urban-Cooling-Data.git. The ambient temperatures and CDH dataset will be accessible after publication both on GitHub and on the website of the Chair of Building Physics, ETH Zurich.

**Appendix A. CDH description and discussion**

CDH is used to quantify the amount of cooling required to maintain indoor comfort in indoor environment during hot weather conditions. CDH is calculated based on the air temperature data, by quantifying what degree and for how long the outdoor air temperature is higher than a base temperature with a resolution of one hour.

The mathematical expression is defined by ASHRAE and is explained in Equation 1 (ASHRAE 2005).

$$\text{CDH} = (1 \text{ hour}) \sum_{\text{hours}} (t_{oa} - t_b)^+ \qquad (1)$$

where $t_b$ is the base temperature and $t_{oa}$ is outdoor ambient temperature at 2m height for every hour. The positive sign (+) after the parenthesis means that only positive values are counted.

The base temperature is typically chosen as a setpoint temperature in the air conditioning system, above which the air conditioning system is operated. It is a metric that employs outdoor air temperature data at 2m height.

Compared to many building energy models, the CDH calculation has its limitations as it does not consider specific building characteristics, for instance the occupancy, building materials, indoor and outdoor humidity, and configurations of cooling and heating systems. Studies that focus on a understanding of building-specific cooling energy demand may integrate CDH with building-scale weighting parameters (Spinoni *et al* 2018, 2015). In our work, to examine three-decade, city-scale evolution in cooling energy due to global warming, urban heat islands and heat extremes, the CDH is a simple yet representative metric.

**Appendix B. Thermal comfort analysis according to indoor air conditions**

An understanding of the adaptive thermal comfort in the local context is required to make decisions on setpoint parameters for cooling, which determines a building's cooling load and cooling efficiency. In addition to the setpoint air temperature, setpoint relative humidity also affects the thermal comfort in an indoor environment. To quantify thermal comfort in an indoor setting, we use Standard Effective Temperature (SET) with several basic assumptions in which the mean radiant temperature is equal to the air temperature, an air speed of 0.1 m/s, a metabolic rate of 1 met, and a clothing level of 0.6 clo representing summer clothing conditions (Ji *et al* 2022). Table B shows the SET values of an indoor setting with air temperature in the range of 22-27 °C and relative humidity in the range of 30% to 70%.

Table B. The Standard Effective Temperature (SET, °C) calculation metrics based on a range of indoor air temperatures and relative humidity levels. The color represents human sensation and physiology of SET values.

| SET (°C) | | Indoor Air Temperature (°C) | | | | | |
|---|---|---|---|---|---|---|---|
| | | 22 | 23 | 24 | 25 | 26 | 27 |
| Indoor Relative Humidity | 30% | 21.4 | 22.4 | 23.4 | 24.3 | 25.2 | 26.1 |
| | 40% | 21.5 | 22.5 | 23.5 | 24.5 | 25.5 | 26.4 |
| | 50% | 21.6 | 22.6 | 23.7 | 24.7 | 25.7 | 26.8 |
| | 60% | 21.7 | 22.8 | 23.8 | 25.0 | 26.1 | 27.2 |
| | 70% | 21.8 | 22.9 | 24.0 | 25.3 | 26.5 | 27.8 |

| SET (°C) | Sensation | Physiology |
|---|---|---|
| 25.6-30 | Slightly warm, slightly unacceptable | Slight sweat, vasodilation |
| 22.2-25.6 | Comfortable, acceptable | Physiological thermal neutrality |
| 17.5-22.2 | Slightly cool, slightly unacceptable | Initial vasoconstriction |